\documentclass[11pt,twoside]{article}

\usepackage{asp2006}
\usepackage{epsfig}
\usepackage{graphicx}

\def\media#1{\langle #1\rangle}

\begin{document}
\title{On the Origin of Early-type Galaxies Nuclei}   
\author{R. Capuzzo Dolcetta$^{(1)}$, P. Miocchi$^{(1)}$}   
\affil{$^{(1)}$ Dep. of Physics, Universit\'a di Roma ``La Sapienza'' , Roma, Italia}    

\begin{abstract} 
The ACS Virgo cluster survey by C\^ot\'e and collaborators shows the presence of compact nuclei at the
photocenters of many early-type galaxies. 
It is argued that they are the low-mass counterparts of nuclei hosting Super Massive Black 
Holes (SBHs) detected in the bright galaxies. If this view is correct, then one should 
think in terms of central massive objects, either SBHs or Compact Stellar Clusters (CSCs), 
that accompany the formation of almost all early-type galaxies. In this observational 
frame, the hypothesis that galactic nuclei may be the remains of globular clusters driven inward to the galactic 
center by dynamical friction and there merged, finds an exciting possible confirm. 
In this short paper we report of our recent results on globular cluster mergers obtained
by mean of detailed $N$-body simulations.  
\end{abstract}


\section*{Globular cluster merging in the galactic centers}   

The Virgo Cluster Survey performed by \citet{cote06}, by mean of 
the ACS on the Hubble Space Telescope, has given evidence of
the presence of compact nuclei at the photocenters of many
early-type galaxies in the cluster, whose luminosity distribution is much
better fitted by an extended (King's) profile rather than by a point source
\citep[see][for a similar finding in late-type spirals]{bok02}. 
The half mass radii of nuclei, $r_h$, are in the range  $2<r_h$[pc]$<62$, 
with $\media{r_h} = 4.2$ pc, and correlate with the nucleus
luminosity: $r_h \propto L_n^{0.5\pm 0.03}$.
The mean of the frequency function for the nucleus-to-galaxy luminosity ratio in
nucleated galaxies, $\log\eta = - 2.49 \pm 0.09$, is indistinguishable from
that of the SBH-to-bulge mass ratio,
$\log (M_{BH}/M_{gal}) = -2.61 \pm 0.07$, calculated in 23 early-type galaxies.\\ 
On these bases, \citet{cote06} argue that resolved stellar nuclei are 
the low-mass counterparts of nuclei hosting SBHs  detected in the bright
galaxies. 
If this view is correct, then one should think in terms of central massive
objects, either SBHs or CSCs, that accompany the 
formation and/or early evolution of almost all early-type galaxies.

It is clear that these characteristics of early-type galaxies nuclei well fit
into a scenario of multiple globular cluster (GC) merging in the inner galactic regions, 
corresponding to a ``dissipationless'' scenario as alternative to the ``dissipational''
scenario, this latter being based mainly on speculative hypotheses
\citep[see, e.g.,][]{babrees92} supported just by some quantitative
results \citep{mihhernq94}.
The validity of the ``dissipationless'' hypothesis requires, at first, a detailed
$N$-body simulation of the interaction of stellar clusters in the inner regions of
their parent galaxy, keeping into account both the cluster-cluster and the
cluster-galaxy dynamical interaction. This latter depends on tidal distortion,
acting on the cluster internal motion, and on dynamical friction, acting on the 
cluster orbital motion.
Not many simulations have been presented in the literature that study the
possible formation scenarios for Nuclear Clusters (NCs). Among these, we remind those by
\citet{fell05} finding that super-massive star clusters, like W3 in NGC 7252, are
very likely the merging remnants of objects spanning
a range of masses from $\omega$ Cen scale to that
of ultra-compact dwarf galaxies up to very small dwarf ellipticals. 
Interestingly enough, the resulting super cluster is a stable and bound
object, whose density profile is well fitted by a King profile, even if 
mass loss of the merger product occurs through every perigalacticon passage.
Also \citet{bekki04} examined a merger formation scenario
in which NCs are the product of multiple merging of star
clusters in the central regions of galaxies. \citeauthor{bekki04} simulations
are done in a simplified way, neglecting the role of the galactic external
potential and with a relatively low resolution. Therefore, it is not clear
how realistic are their interesting results showing (i) the full merging of the 
interacting clusters in a significantly short time ($\sim 10^7$ yr), (ii) the
triaxiality in the merger configuration and (iii) the existence of scaling relations
among velocity  dispersion, luminosity, effective radius, etc..

\section*{The Simulations}   
Here we report briefly of some of the results presented in \cite{cdmioc07}
 where we studied whether and how  the merging of various massive GCs decayed by dynamical friction in the inner galactic region may occur.\\
The main questions to answer are: (i) given some (realistic) initial 
conditions for a set of GCs which experienced a significant orbital decay, 
are they undergoing a  full merge? (ii) if so, what is the time needed? 
(iii) what is the final structure  of the merged supercluster?
(iv) does it attain a quasi-steady state?

The answers to these question are of overwhelming importance to give 
substance to the interpretation of the formation of early type galaxy nuclei via
merging of decayed GCs, a hypothesis raised first by \citet{tremetal75} and
subsequently extensively studied by Capuzzo-Dolcetta and collaborators 
(see \citet{cv05} and references therein). 
We consider GCs as $N$-body systems moving within a triaxial
galaxy represented by an analytical potential, subjected also to the deceleration due
to dynamical friction. 
We studied the merging process occurring among four GCs already decayed in the
inner region of the galaxy. The galaxy where the GCs move is represented by a
self-consistent triaxial potential. 
In this potential, the dynamical friction yields decaying times significantly
shorter than the Hubble time \citep{cv05}. 
We have simulated 4 massive GCs having an initial King profile with total 
mass, central velocity dispersion and core radius ranging in
$M = 42$ -- $54 \times 10^6$ M$_\odot$, $\sigma_0 = 25$ -- $36$ km s$^{-1}$,
$r_c = 2.2$ -- $3.8$ pc. They are initially located within $100$ pc from the
galactic center (Fig. 1). Simulations are done with the parallel `ATD' $N$-body code
\citep{miodol02}, with a total of $10^6$ particles.  

\section*{Results}
The merging occurs rather quickly: after $\sim 15$ Myr, corresponding to
$\sim 20$ galactic core crossing times ($t_{cross}$), the merging is completed and the
resulting system attains a quasi-equilibrium configuration (see Fig. 1).
The total projected density profile in the nuclear region is remarkably 
similar to those recently observed in the central regions of nucleated 
early-type galaxies in the Virgo cluster (Fig. 2). 
The final CSC morphology is that of an axisymmetric ellipsoid (axial 
ratios $1.4$:$1.4$:$1$, ellipticity $\sim 0.3$) without figure rotation.
In the velocity dispersion-mass plane, the CSC is located closer to the scaling relation followed 
by GCs than to that of elliptical galaxies. This is due to that
our system  is located deep inside the galaxy potential.

\begin{figure*}
\centering
\includegraphics[width=10.0cm]{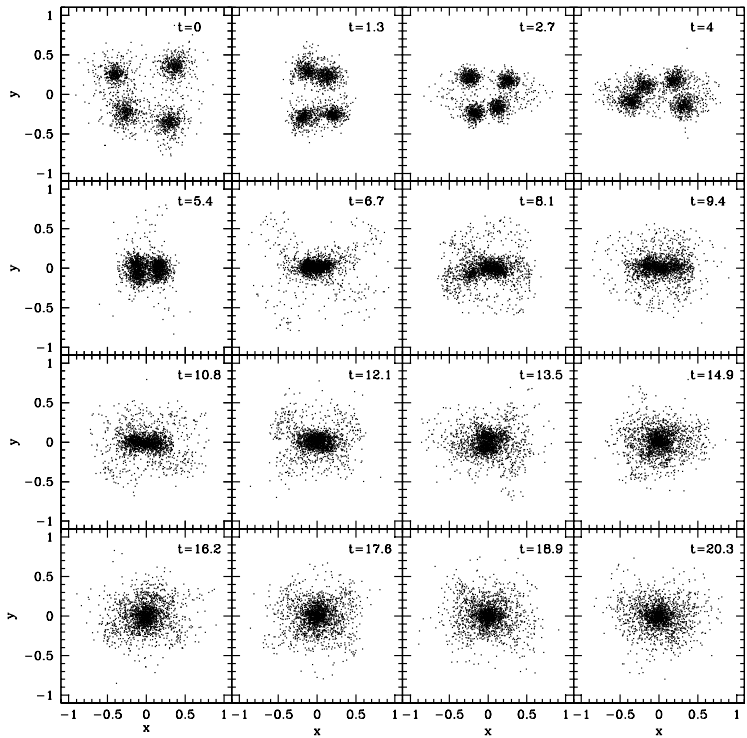}
\caption{Snapshots of the clusters evolution (projected on the $x$--$y$ plane)
from $t=0$ to $t=20.3$ Myr. Lengths are in units of $200$ pc and time is in
Myr.}
\label{snap1}
\end{figure*}

  \begin{figure*}
   \includegraphics[width=6.2cm]{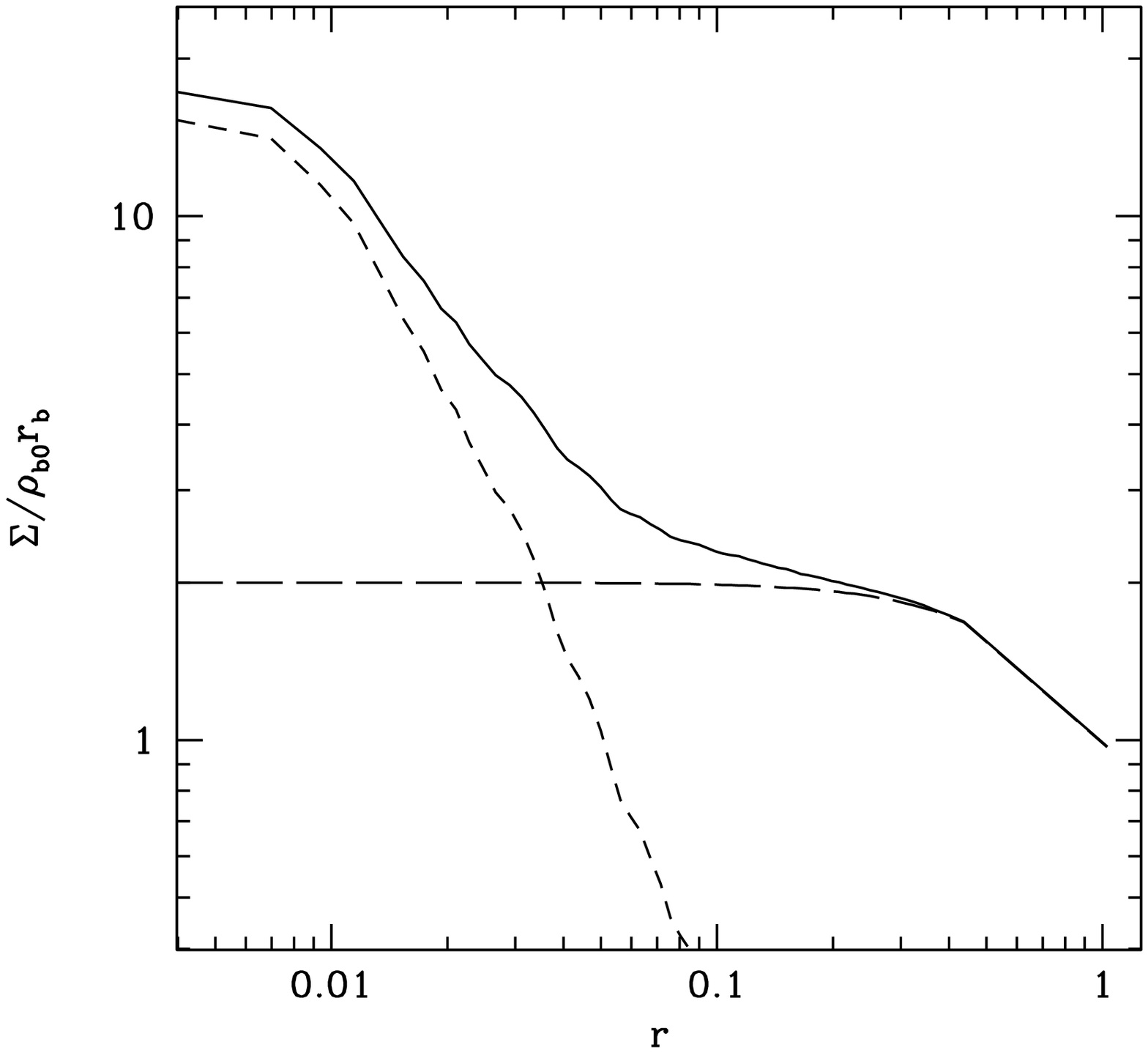}
   \includegraphics[width=6.2cm,height=6cm]
{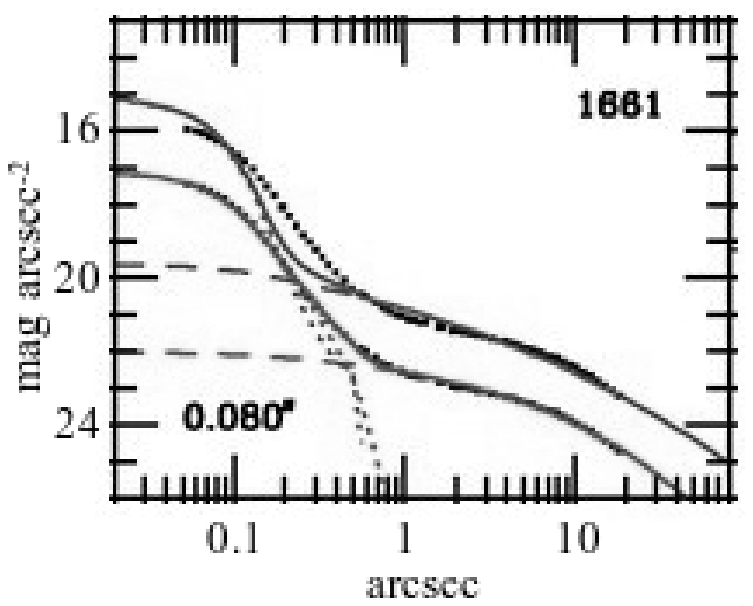}
\caption{Left panel: Surface density profile (in proper units) for our last computed
CSC configuration (short dashed line), overlapped to the galaxy profile 
(long dashed) so to give the total density (solid line).
Right panel: surface brightness profile
of NGC 1661, 
where the lower solid line is the King's model fitting to data (black dots) 
and the upper solid line is the point source fitting. Dashed line is the
galaxy contribution. All the curves are translated for display convenience.
The median half light radius
($0.080"$) is reported. \citep[from][]{cote06}.}  
        \label{fig1}
    \end{figure*}

\section*{Conclusions}
We studied the modes of interaction of a few (4) globular clusters
which, due to their large initial masses, have experienced a significant
dissipation of their orbital energy by dynamical friction caused by the galactic triaxial
field. The GC motion in the inner galactic region is characterized by a further
barycenter braking due to the galactic tidal torque, followed by a full merge that
is completed in $\sim 20$ $t_{cross}$.
The merger result keeps some of the characteristics of the preexisting objects,
leading to a single cluster with a King density profile plus a halo which is diffusing
in the external field.
Due to this, we can speak with some confidence of a secularly stable Compact Star Cluster,
Even in the limits of our simulations (both on the number of merging objects and on the 
integration time, which covers $\sim 45$ $t_{cross}$ ), we may extrapolate from our results
that the nuclei of many early-type elliptical galaxies
 may have formed by the merging of few tens of GCs massive enough to have decayed
into the inner galactic regions in a time short 
compared to the Hubble time and compact enough to survive to the tidal stresses.



\begin{thebibliography}{}
\bibitem[Babul \& Rees, 1992]{babrees92}Babul, A., Rees, M.J. 1992, MNRAS, 255, 346
\bibitem[Bekki et al., 2004]{bekki04}K. Bekki, W.J. Couch, M.J. Drinkwater, Y. Shioya:
ApJ 610, L13 (2004)
\bibitem[B\"oker et al., 2002]{bok02}
B\"oker, T., Laine, S., van der Marel, R. P., Sarzi, M., Rix, H.-W., Ho, L. C.,
 Shields, J. C. 2002, AJ, 123, 1389
\bibitem[Capuzzo-Dolcetta \& Miocchi, 2007]{cdmioc07}R. Capuzzo Dolcetta, P. Miocchi: in preparation
(2007) 
\bibitem[Capuzzo-Dolcetta \& Vicari, 2005]{cv05}R. Capuzzo-Dolcetta, A. Vicari: MNRAS 356, 899 (2005)
\bibitem[C\^ot\'e et al., 2006]{cote06}P. C\^ot\'e et al.: ApJSS 165, 57 (2006)
\bibitem[Fellhauer \& Kroupa, 2005]{fell05} M. Fellhauer, P. Kroupa: MNRAS 359, 223 (2005)
\bibitem[Mihos \& Hernquist, 1994]{mihhernq94} Mihos, J.C., Hernquist, L.
1994, ApJ, 437, L47
\bibitem[Miocchi \& Capuzzo-Dolcetta, 2002]{miodol02}P. Miocchi, R. Capuzzo Dolcetta: A\&A 382, 758 (2002)  
\bibitem[Tremaine et al., 1975]{tremetal75}Tremaine, S., Ostriker, J.P., Spitzer,
L. Jr. 1975, ApJ, 196, 407
\end{thebibliography}
\end{document}